\pdfoutput=1
\documentclass[
aps,reprint,nobibnotes,superscriptaddress,prl]{revtex4-2}

\usepackage{float}
\usepackage[utf8]{inputenc}
\usepackage{orcidlink}
\usepackage{graphicx,amsmath,amsfonts,amssymb,slashed}
\usepackage{bbold,wasysym}
\usepackage{enumitem,natbib}

\usepackage{bbm,euscript,hyperref}
\hypersetup{colorlinks, citecolor=blue, linkcolor=newred, urlcolor=blue}
\usepackage{color}

\usepackage{graphicx,capt-of}

 \definecolor{newred}{rgb}{0.75, 0., 0.25}

\begin{document}

\title{Nonlinear Gravitational Wave Memory : Universal Low-Frequency Background
}
\author{Caner \"Unal \orcidlink{0000-0001-8800-0192}} 
\email{unalc@metu.edu.tr}
\email{caner.unal@nanograv.org}

\affiliation{Fizik B\"ol\"um\"u, Orta Do\u{g}u Teknik \"Universitesi, \c{C}ankaya, Ankara 06800, Türkiye}

\author{Do\u{g}a Veske \orcidlink{0000-0003-4225-0895}} 
\email{veske@metu.edu.tr}
\affiliation{Fizik B\"ol\"um\"u, Orta Do\u{g}u Teknik \"Universitesi, \c{C}ankaya, Ankara 06800, Türkiye}
\affiliation{Columbia Astrophysics Laboratory, Columbia University, New York, NY 10027, USA}

\begin{abstract} 
A universal contribution exists in the infrared (low frequency) regime of all gravitational waves, which results from nonlinear memory. Nonlinear memory is sourced by linear order gravitational waves and exists for any gravitational-wave background. We calculate the stochastic nonlinear memory signal of various stochastic backgrounds of cosmological (scalar induced, reheating, phase transition, topological defect, turbulence) and astrophysical (binary mergers of stellar-mass, intermediate mass, supermassive, and primordial black holes) origins. These results allow us to derive the complete frequency spectrum of cosmological and astrophysical SGWB. We calculate how to probe the thermal state of the universe, i.e. the equation of the state, via the memory spectrum's slope and also discuss the detection prospects at various frequency bands with future experiments.
\end{abstract}

\maketitle
\noindent

{\it Introduction---}
Gravitational waves (GW) are propagating degrees of freedom of the geometry and preserve valuable information about the processes that generate them. Various astrophysical and cosmological processes can lead to the production of GW. By characterizing the properties of GW, one can learn about the relevant processes and fundamental physics.  The current and future experimental efforts encompass different detectors, ground and space based, from table-top size to galactic or even observable-universe size, to investigate GWs over approximately 30 decades of frequency range, i.e. $10^{-19} - 10^{11}$~Hz, including current horizon scale frequency band~\cite{BICEP:2021xfz,LiteBIRD:2022cnt}, nHz band~\cite{NANOGrav:2023gor,NANOGrav:2023hvm,EPTA:2023fyk,EPTA:2023xxk,Reardon:2023gzh,Xu:2023wog,Janssen:2014dka,Weltman:2018zrl}, $\mu$Hz band~\cite{Sesana:2019vho,Foster:2025nzf}, mHz band~\cite{LISA:2022yao,LISACosmologyWorkingGroup:2022jok,LISA:2022kgy,TianQin:2015yph,Hu:2017mde,Ruan:2018tsw,Alves:2024ulc}, deciHz band~\cite{Yagi:2011wg,Alves:2024ulc}, Hz band~\cite{LIGOScientific:2014pky,LIGOScientific:2016wof,Punturo:2010zz,Acernese_2014,2021PTEP.2021eA101A}, kHz and above band~\cite{Aggarwal:2020olq,Aggarwal:2025noe}.

Gravitational field couples to energy and momentum. Moreover, gravitational field itself possesses energy and momentum. Hence, gravitational field couples to itself. This induces an inherent nonlinearity in the gravitational interactions~\cite{Zeldovich:1974gvh,Smarr:1977fy,Christodoulou:1991cr,
Blanchet:1992br,PhysRevD.45.520,Will:1996zj,Favata:2010zu,Strominger:2014pwa}. One direct consequence of this self-coupling is the nonlinear GW memory. The general GW memory effect is a persistent, non-oscillatory displacement in the spacetime metric induced by the permanent loss of energy from the GW-emitting source~\cite{1987Natur.327..123B} (for cosmological background \cite{Tolish:2016ggo,Bieri:2017vni}). Generally, this energy loss is sourced by any emission process of at least quadrupolar order. The nonlinear GW memory is a special case of it which is sourced by the energy-momentum carried away by the emitted GWs themselves. Therefore, nonlinear memory is universally produced by all GW sources. Thanks to this clear prediction, the nonlinear memory can be a clean probe of gravitation theory and high-energy theories~\cite{Pate:2017fgt,Heisenberg:2023prj}; and there is experimental effort to detect it \cite{Lasky_2016,NANOGrav:2023vfo,Gasparotto:2023fcg,Cheung_2024,Inchauspe:2024ibs,Agazie:2025oug,Gasparotto:2025wok}.

There are numerous types of GW sources with distinct GW properties (amplitude, peak frequency, shape, statistical distribution). Loud sources are rarer and distinguishable, whereas the unresolvable signals from the fainter sources produce an effectively stochastic overall signal. The claimed 3$\sigma$ detection of the nHz GW signal with pulsar timing arrays~\cite{NANOGrav:2023gor}, or the expected signal from galactic double white dwarfs~\cite{Ruiter_2010} which may be a noise source for mHz GW detectors are such signals. The GW memory signal is no exception for creating a stochastic signal. There should be a stochastic signal composed of unresolved memory signals from astrophysical and cosmological sources.

We investigate the formal, phenomenological and observational aspects of stochastic nonlinear memory background. We first calculate the stochastic nonlinear memory backgrounds in a radiation dominated universe, and derive complete nonlinear memory solution for both sub-horizon and super-horizon scales. Then, we generalize our results for a generic equation of state and expansion rate, which allows us to probe the thermal state of the universe via memory spectrum's slope. We also explore the regimes where memory dominates over linear signal. We predict nonlinear memory backgrounds for a wide spectrum of cosmological and astrophysical GW backgrounds, allowing us to derive complete frequency spectrum. We finally discuss the observational aspects of the nonlinear memory background and show that due to slower decay of the memory signal's spectrum at low frequencies, it might allow the detection of a single phenomenon by multiple detectors at multiple frequency bands via future sensitive detectors at different frequencies.

{\it Stochastic Gravitational Wave Backgrounds---}
In the FLRW background with conformal time and space ($\eta$, {\bf x} respectively) coordinates, and speed of light $c=1$ such that the differential spacetime distance is  $ds^2 = a^2(\eta) ( - d \eta^2 + d {\bf x}^2)$, and the equation for GW evolution in transverse traceless (TT) gauge is given by
\begin{equation}
h^{''}_{i j} + 2 \mathcal{H} h_{i j}' + k^2 h_{i j} = a^2 \, 16 \, \pi \, G_{\rm N} \, T_{ij}^{\rm TT} =
 {\cal S}_{i j}^{\rm TT} 
\end{equation}
where prime denotes conformal time derivative, $\mathcal{H}= a'/a$ is the conformal Hubble parameter, $k$ is the conformal wavenumber, $G_{\rm N}$ is the gravitational constant, and $a$ is the scale factor.

The fraction of energy density in GW per logarithmic frequency bin is given by
\begin{equation}
  \Omega_{\mathrm{GW}} = \frac{1}{\rho_c} \frac{\mathrm{d} \rho_{\mathrm{GW}}}{\mathrm{d} \log k} = \frac{ {\overline{ \langle h' h'\rangle}}}{12 H^2 a^2} = \frac{1}{24} \frac{k^2}{a^2 H^2} \; \underset{\lambda}{\Sigma} \; \Delta_{h, \, \lambda}^2 (k)
  \label{eqgwenergydensity}
\end{equation}
where $\rho_{\rm GW}$ is the energy density of GW, $\rho_c=\frac{3\mathcal{H}^2}{8\pi G}$ is the critical energy density of the universe, overline indicates averaging over multiple periods, $\lambda$ represents the polarizations of GW. The two-point function is given by
\begin{equation}
  \langle h_{{\bf k}, \, \lambda} \; h_{{\bf k'}, \, \lambda'} \rangle 
  =  \frac{2\pi^2}{k^3} \; \Delta^2_{h,\lambda}(k) \; \delta_{\lambda \lambda'} \; \delta^3({\bf k}+ {\bf k'}) .
  \label{eqhtwopoint}
\end{equation}

{\it Stochastic Nonlinear GW Memory Background---} The nonlinear GW memory is sourced from the causal (sub-horizon) linear order GWs. Its evolution is given by
\begin{equation}
h^{''}_{i j} + 2 \mathcal{H} h_{i j}' + k^2 h_{i j}
 = {\cal S}_{i j}^{\rm TT} \simeq  \frac{1}{2} \partial_i h_{kl}^{\rm TT} \partial_j h_{kl}^{\rm TT}
\end{equation}
and we drop TT superscript for the rest.

We extract out polarizations of the GW by using projection tensors, $\epsilon_{ij}^{\lambda}(k)= \epsilon_{i}^{\lambda}(k) \epsilon_{j}^{\lambda}(k)$ where $\lambda$ being two polarizations \footnote{They satisfy relations $\epsilon_{i}^{\lambda}(-{\bf k})=\epsilon_{i}^{*\lambda}({\bf k})$,  $\epsilon_{i}^{\lambda}({\bf k}) \, \epsilon_{i}^{*\lambda}({\bf k}) =1$, $\epsilon_{i}^{\lambda}(k) k^i =0$, $\vert \epsilon_{i}^{\lambda}(k) p^i \vert^2 = \frac{1}{4} \left( 1- \lambda \; {\hat p}\cdot {\hat k} \right)^2$.}. The generic solution is given by
\begin{equation}
  h_{\lambda} ({\bf k}, \eta) = \int^{\eta}_{\eta_*} d\eta' \; \frac{a(\eta')}{a(\eta)} \; G ({\bf k},\eta,\eta') \; \epsilon_{ij}^{\lambda}(k) \; {\cal S}^{ij}({\bf k},\eta')
  \label{eqhparticular}
\end{equation}
where $G ({\bf k}, \eta, \eta')$ is the Green function.

To find $\Omega_{\mathrm{GW}}$, we use Eq. \eqref{eqhparticular} in Eq. \eqref{eqhtwopoint}, and plug to Eq. \eqref{eqgwenergydensity}.
We first assume a radiation dominated universe, later we generalize our results to a generic equation of state. In radiation domination, we have $a\propto \eta$ and $G ({\bf k},\eta,\eta')=\sin [k (\eta - \eta')]/k$. 
We find the energy density parameter of the nonlinear memory as
 \begin{multline}
   \Omega_{\rm GW} = \frac{k^3}{24 \, \pi^2 \, a^2 H^2} \int^{\eta}_{\eta_*} \int^{\eta}_{\eta_*} d  \eta' \; d \eta'' \frac{\eta' \, \eta''}{\eta^2} \,  \; \\
   \times k \, G( {\bf k}, \eta,\eta') \; k\, G ( {\bf k'},\eta,\eta'') \; P_{\cal S}(k)
   \label{eqomeganonlinearmemory}
 \end{multline}

where the source two-point function is $ \langle {\cal S}_k {\cal S}_{k'} \rangle = \delta^3 ({\bf k} + {\bf k'}) \; P_{\cal S}(k) $. Our source is given as
 \begin{eqnarray}
\!\!\! \!\!\! P_{\cal S}(k) &=& \int \int d^3p \; d^3q \; p \, \vert k-p\vert \, q \, \vert k'-q\vert \, \sin^2{\theta}_{{\bf k,p}} \, \sin^2{\theta}_{{\bf k',q}} \nonumber\\
  &&  \;\;\;\;\; \;\; \times \; \;\;\; \langle h_p^\lambda  h_{k-p}^\lambda h_q^{\lambda'}  h_{k'-q}^{\lambda'} \rangle_\delta \nonumber\\  && \!\!\!\!\!\!\!\!\!\!\!\!\!\!\! \simeq 2 \int d^3p \, p^2 \vert k-p\vert^2 \, \sin^4 \theta_{\bf k,p} \, \frac{4\pi^4}{p^3 \vert {\bf k}- {\bf p} \vert^3} \Delta_h^2(p) \Delta_h^2(k-p) \nonumber\\
  && \hspace{-0.5cm} \simeq 2\pi^2 \, \frac{16}{15} \; k_* \; {\cal F}\; \Delta_h^4 (k_*)
  \label{eqmemorysourcedimensionless}
 \end{eqnarray}
where we evaluate the four-point correlator via Wick theorem using Eq. \eqref{eqhtwopoint} assuming Gaussian fields \footnote{In principle non-Gaussian properties can introduce ${\cal O}$(a few) corrections and usually h is non-Gaussian to some extent which will be subject of future investigation.}, $k_*$ is the GW spectrum peak momentum, $\theta_{\bf k,p}$ is the angle between $\bf{k}$ and $\bf{p}$, and we took the $k\ll p$ limit in the last line. The factor of 2 comes from two possible contractions, solid angle integral gives $\int \frac{d\Omega}{4\pi} \sin^{4} \theta = \frac{8}{15} $, and $\langle \rangle_\delta$ denotes removing momentum conserving Dirac-delta $\delta^3 (k+k')$ from the correlator. ${\cal F}$, a dimensionless factor, depends on the spectrum shape, given by 
\begin{equation}
  \int \frac{dp}{k_*} \; \frac{\Delta_h^4 (p)}{\Delta_h^4(k_*)}.
\end{equation}

We plug Eq. \eqref{eqmemorysourcedimensionless} into Eq. \eqref{eqomeganonlinearmemory} and compute the time integrals, whose details are given in the End Matter. The nonlinear GW memory spectrum in the $k<k_*$ regime is given as
\begin{equation}
\Omega_{\rm GW, \, mem} \simeq {\cal A} \; \frac{k}{k_*} \;  \bigg[1+\ln^2\left(\frac{ 1+ k \eta_*}{k \eta_*}\right)\bigg] \bigg[\frac{k \eta_*}{1+k \eta_*}\bigg]^2
  \label{eq:genericformemory}
\end{equation}
with $k\eta_*=k/{\cal H}_*$, and ${\cal H}_*$ being the conformal Hubble
\begin{equation}
  {\cal A} \equiv \frac{16}{15} \; {\cal F} \;  Q\;\Omega_{\rm GW, lin}(k_*) \;  \Delta_h^2 (k_*) \,. 
\end{equation} 
$Q= ( a^4_{k} \; \rho_k) / (a_0^4 \;\rho_0)$ is the scaling of the energy density from the time the mode $``k"$ enters the horizon to now. For radiation domination, $Q= \Omega_{rad,0} \simeq 8\cdot 10^{-5}$.

If it is a sub-horizon scale mode during GW production ($k > {\cal H}_*$), then we have 
\begin{equation}
   \Omega_{\rm GW, \, mem} (k > {\cal H}_*) \simeq {\cal A} \; \frac{k}{k_*}.
   \label{eqsubhorizonuniversal}
\end{equation}

If the mode is super-horizon scale during GW production ($k<{\cal H}_*$), then we have 
\begin{multline}
  \Omega_{\rm GW, \, mem} (k < {\cal H}_*) \simeq {\cal A} \; \frac{k}{k_*} \; \frac{k^2}{{\cal H}_*^2} \left( 1 + \ln^2(k\eta_*) \right)
\end{multline}
As a check, we give the heuristic derivation of the spectrum : In sub-horizon regime, we have a non-expanding background ($a\simeq {\rm const.}$), hence $\int d\eta \sim 1/k$ and overall $\Omega_{\rm GW, \, mem} \propto k$. In super-horizon regime, we have $a\propto \eta$, hence $\int d\eta / \eta \sim \ln 1/k \eta_*$ and overall $\Omega_{\rm GW, \, mem} \propto k^3 \ln^2 1/k\eta_*$ with $k = 2 \pi f$.

\begin{figure*}
  \centering
  \includegraphics[width= 0.49 \linewidth]{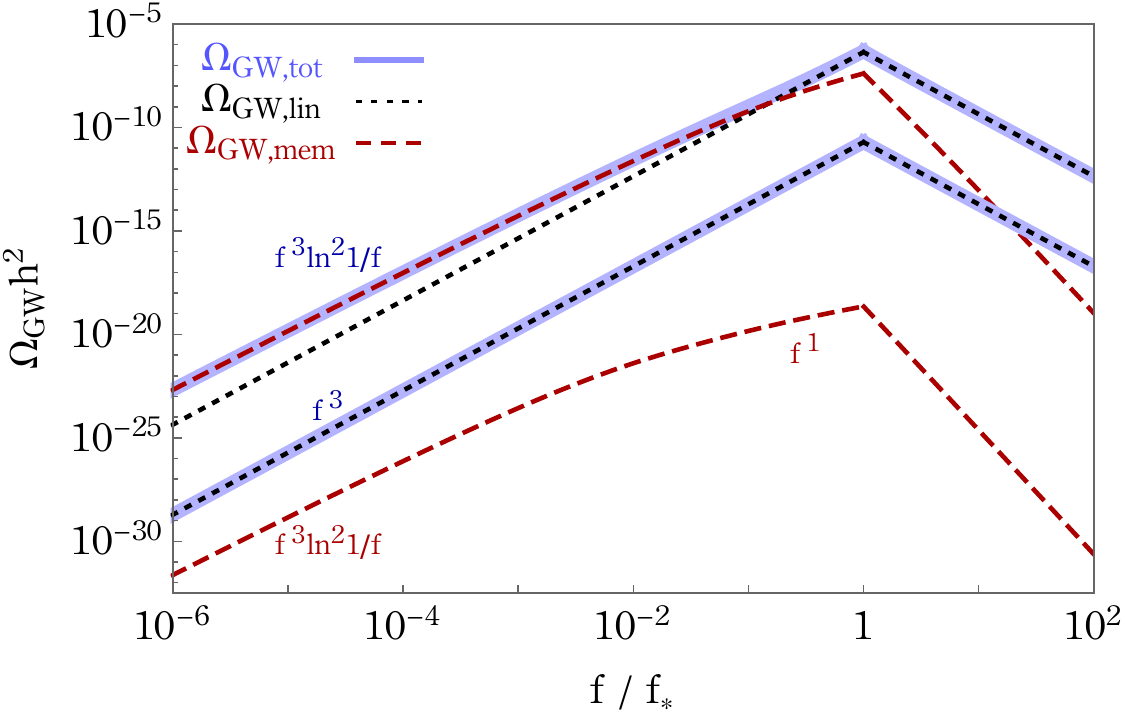}
  \includegraphics[width= 0.49 \linewidth]{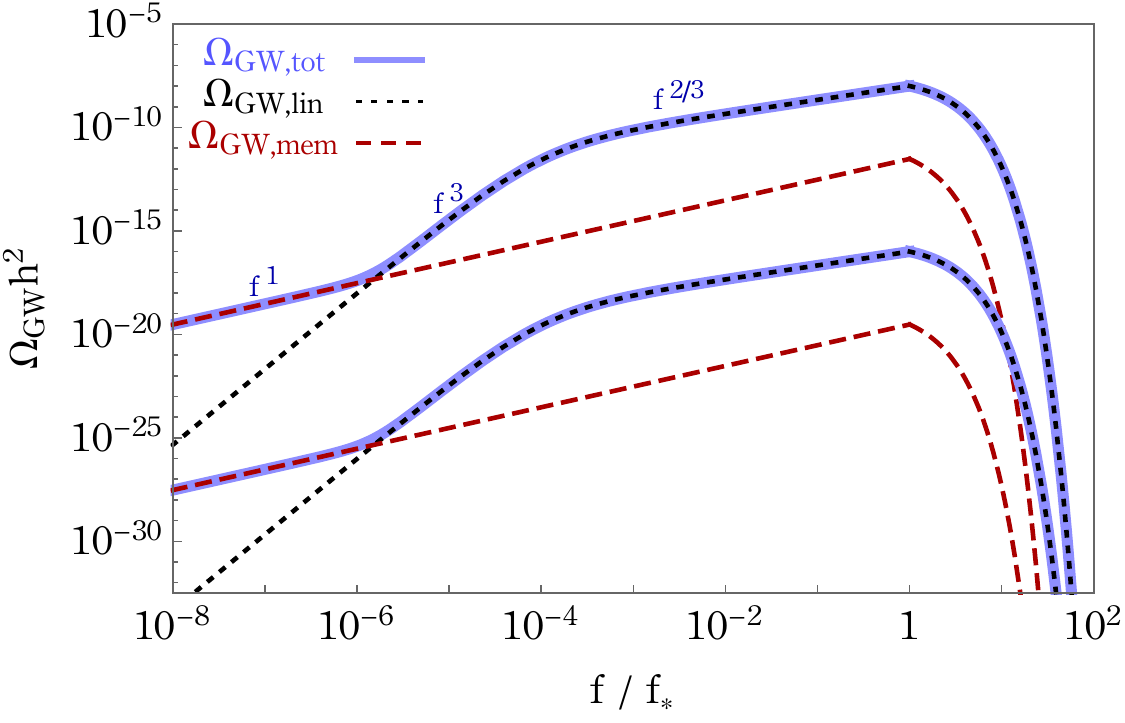}
  
  \caption{Decomposition of GW signal : Linear (black), nonlinear memory (red) and total (blue).
  Left: 
    Character of a cosmic SGWB produced in radiation domination, where peak frequency $f_{*}=k_*/2\pi$ can be anywhere from nHz (MeV) to THz ($10^{18}$ GeV). We choose two sets of example signals to compare the effects of amplitude of linear power spectrum, $\Delta_h^2$, and momentum to Hubble ratio, $k_*/{\cal H}_*$: i)Top one has $k_*/{\cal H}_*\sim 1$ and $\Delta_h^2 (k_*) = 1/4$; ii)The bottom one has $k_*/{\cal H}_*=30$ and $\Delta_h^2 (k_*) = 10^{-8}$. The memory surpasses linear signal at $f^{\rm cosmo}_{\rm mem. dom.}$ according to Eq. \eqref{eqcosmomemdom} or Eq. \eqref{eqgenericcosmomemdom} depending on the equation of state. 
  Right:
  Character of astrophysical SGWBs, where peak frequency $f_{*}$ can be $ 10^3, \, 10^{-1}, 10^{-7}\, {\rm Hz}$ for the SGWB of stellar-mass, intermediate and supermassive BBH mergers. Ratio of linear signal to nonlinear memory is nearly the same for BBH mergers for different masses and event rates, i.e. $ \Omega_{\rm GW,mem} / \Omega_{\rm GW,lin} \approx 2-3 \cdot 10^{-4}$. Below peak frequency, linear signal scales with $f^{2/3}$ until the turn-around, and with $f^3$ below turn-around; while memory scales with $f$. The memory starts dominating the linear signal at $f^{\rm astro}_{\rm mem. dom.}$ according to Eq. \eqref{eqastromemdom}. 
  }
  \label{fig:decomposenonlinearmemory}
\end{figure*}

The spectrum in the sub-horizon ($k>\mathcal{H}_*$) can also be derived as follows: The memory signal behaves like a step function that has a non-zero rise time \footnote{There is also a decay with the timescale of the light travel time from the source; but that corresponds to extremely low frequencies for feasible scenarios and is not our discussion here.}. The power spectral density (PSD), denoted as $S(f)$, of a stochastic background arising from independent memory signals can be found using the fact that two-sided PSD of independent impulses is a white spectrum with an amplitude equal to the variance of the amplitude of the impulses ($\sigma^2$) times the rate of impulses ($\mathcal{R}$). Since the step function is time integral of an impulse,
the PSD of step functions have an extra factor of $(2\pi f)^{-2}$. Consequently, the two-sided PSD of a stochastic memory signal will be $\sigma^2\mathcal{R}(2\pi f)^{-2}$. Using the relation $\Omega(f)=4\pi^2f^3S(f)/(3H_0^2)$~\cite{mingarelli2025understandingomegamathrmgwfgravitationalwave}, we find
\begin{equation}
  \Omega(f)=\sigma^2\mathcal{R}\frac{f}{3H_0^2}.
  \label{eq:omega}
\end{equation} 
This is valid up to the frequency equal to the inverse rise time of the memory signal after which the energy density is lower (see also \cite{Zhao:2021zlr,Boybeyi:2024aax}).

Considering the dominant (2,2) mode contributions to the emitted energy, and the dominant (2,0) mode of the memory signal which is only present in the $+$ polarization in wave coordinates; the memory strain as a function of the inclination angle $\iota$ is given by~\cite{PhysRevD.98.064031,Tolish:2016ggo}
\begin{equation}
  h_{\rm +, wave}^{\rm (2,0), mem.}(\iota)=\frac{4G_{\rm N}(1+z)}{r_{\rm L}c^4}E_{\rm GW}\Gamma\sqrt{\frac{15}{32\pi}}\sin^2(\iota)
\end{equation}
where $\Gamma$ is a factor depending on the multipole structure of the GWs and is about $0.46$ for a linear emission in the (2,2) mode, and $r_{\rm L}$ is the luminosity distance of the source. Considering isotropic source orientations, the average of the observed strain in both polarizations is zero, and the variance is
\begin{equation}
  \sigma^2\left( h_{+,\times,{\rm obs.}}^{\rm (2,0),mem.}\right)
  =\left(\frac{2G_{\rm N}(1+z)}{r_{\rm L}c^4}E_{\rm GW}\Gamma\sqrt{\frac{1}{2\pi}}\right)^2.
\end{equation}
From Eq. \eqref{eq:omega}, the energy density for sources that emit $E_{\rm GW}$ of energy in GW with local rate density $\dot{n}(z)$ is
\begin{equation}
  \Omega(f)=\frac{f}{3H_0^2}\int\left(\frac{2G_{\rm N}(1+z)}{r_{\rm L}c^4\sqrt{2\pi}}E_{\rm GW}\Gamma\right)^2\frac{\dot{n}(z)}{1+z}\frac{dV_c}{dr_{\rm L}}dr_{\rm L}
\end{equation}
where $V_c$ is the comoving volume.

{\it GW Memory in Generic Equation of State---}  
For sub-horizon modes of the memory, we have a universal scaling which is linear in frequency and derived in Eq. \eqref{eqsubhorizonuniversal}, and independent of the background equation of state
\begin{eqnarray}
\Omega_{\rm GW, mem}  &\propto& f.
\end{eqnarray}
Using  Eqs. \eqref{eqhparticular}, \eqref{eqhtwopoint},  \eqref{eqgwenergydensity}, and $a (\eta) \propto \eta^n$ with  $n=2/(3w+1)$ for generic equation of state, we have the scaling of the super-horizon memory modes as
\begin{eqnarray}
  \Omega_{\rm GW, mem}  &\propto& f^5 \left(\int \frac{d\eta'}{a(\eta')} G \right)^2  \propto  f^5 \left( \int \frac{d\eta'}{\eta'^{\,n}} G \right)^2 
  \nonumber\\ 
   &\propto& f^5 \left( \int \frac{d\eta'}{\eta'^{2/(3w+1)}} G \right)^2  \propto f^{3 - 2\, \vert \frac{3w-1}{3w+1} \vert} 
\end{eqnarray}

where $G$ is the Green function, given in the End Matter for a generic background. Notice that if $n>1$, then source is only temporarily on and decays away, hence the main contribution to time integral comes from initial conditions of the source. Since the integral support is from early times, i.e. lower bound, this gives $\Omega_{\rm GW, mem}\propto f^{5-2n} \propto f^{3 + 2\, \frac{3w-1}{3w+1} }  $. This slope is universal for the infrared regimes when the source is active only temporarily, and it is called as causality tail \cite{Caprini:2009fx,Unal:2018yaa,Cai:2019cdl,Hook:2020phx,LISACosmologyWorkingGroup:2024hsc}. On the other hand, if $n<1$, then source is persistent and stays active until times, i.e. $1/k$ and the integral support is from late times, which gives $\Omega_{\rm GW, mem} \propto  f^{1+2n} \propto f^{3 - 2\, \frac{3w-1}{3w+1} } $. We call this scaling as {\it memory tail} due to persistent behavior. This result shows that the slope cannot be steeper than $f^3$, and allows a slower decay for the memory depending on the equation of state. Therefore, the low frequency regime of GWs can be dominated by either causal tail (if linear GW decays away) or memory tail (if linear GW is persistent) depending on the equation of state.

The solution for a generic equation of state becomes
\begin{multline}
   \Omega_{\rm GW, \, mem} \simeq {\cal A} \; \frac{k}{k_*} \; \frac{ (k \eta_*)^2}{1+ 4 (k \eta_*)^2} \\
   \times \left[ \left( \frac{ (k \eta_*)^2}{{\cal C}_1^{\frac{1}{1-n}}+ (k \eta_*)^2} \right)^{n-1} + \left( \frac{ (k \eta_*)^2}{{\cal C}_2^{\frac{1}{n-1}}+ (k \eta_*)^2} \right)^{1-n} \right]
   \label{eqomegagwforgenericeos}
\end{multline}
where ${\cal C}_1\, , {\cal C}_2$ are defined in the End Matter.
The equation above is valid 

for $w\neq1/3$, and for $w=1/3$ see Eq. \eqref{eq:genericformemory}. We see that as $w>1/3$ ($w<1/3$) we have memory (causal) tail and the solution follows the first (second) term in square brackets , i.e. $\Omega _{\rm GW} \propto f^{1+2n}$ ($\Omega _{\rm GW} \propto f^{5-2n}$)\footnote{If there is a background matter that results in $n>5/2$, we see that causal tail has negative slope which implies increasing $\Omega _{\rm GW}$, i.e. larger relative GW energy density compared to critical energy density. This leads to a greater fraction of energy budget of universe in GW, which modifies equation of state of the background with larger radiation content and backreacts on the GW background slope.}.

\begin{figure*}
  \centering
  \includegraphics[width= 0.49 \linewidth]{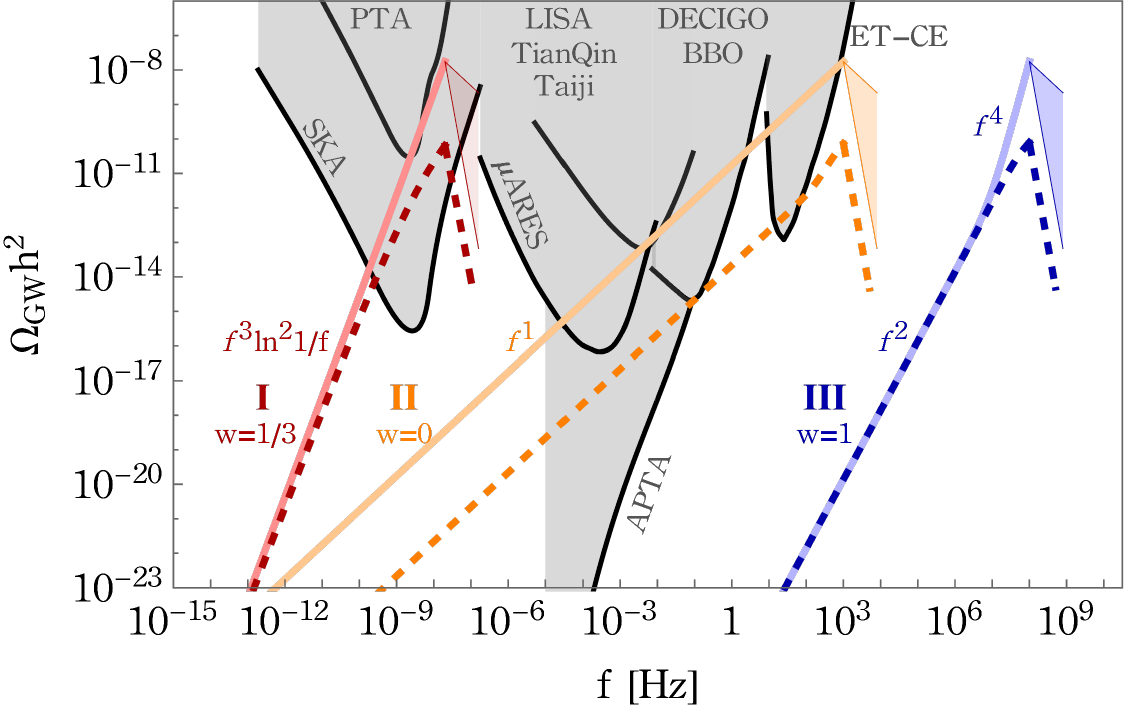}
   \includegraphics[width= 0.49\linewidth]{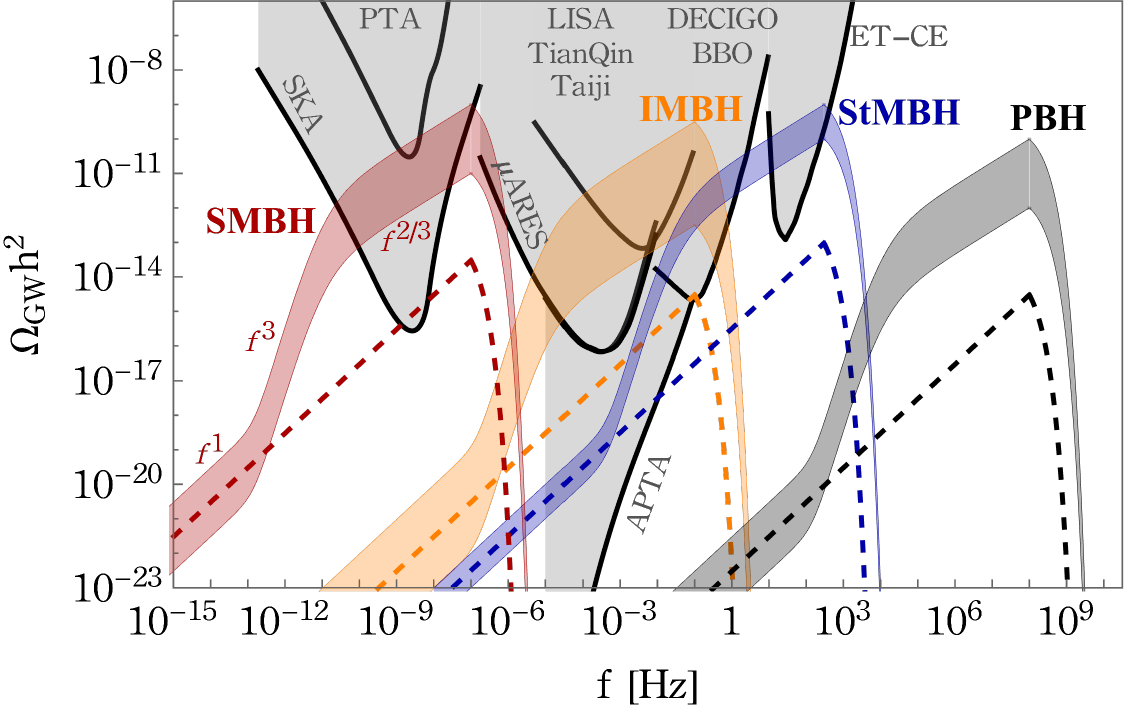}
  \caption{Solid (dashed) lines denote the total (memory) signals. Left: SGWB from cosmic sources with $\Delta_h^2=0.01$ and $k_* \simeq {\cal H}_*$. Memory dominates the linear signal, with a gentle logarithmic dependence for (I) in radiation domination, and with quadratic power for (III) in kination domination. Memory is subdominant for (II) in matter domination. Three example phenomena have peaks at distinct frequencies :  (I) nHz band (PTA~\cite{NANOGrav:2023gor,NANOGrav:2023hvm,EPTA:2023fyk,EPTA:2023xxk,Reardon:2023gzh,Xu:2023wog} and SKA~\cite{Janssen:2014dka,Weltman:2018zrl}),  
(II) mHz (LISA~\cite{LISA:2022yao,LISACosmologyWorkingGroup:2022jok,LISA:2022kgy}, TianQIN~\cite{TianQin:2015yph}, Taiji~\cite{Hu:2017mde,Ruan:2018tsw}, DECIGO-BBO~\cite{Yagi:2011wg}, $\mu$ARES~\cite{Sesana:2019vho} and APTA~\cite{Alves:2024ulc}) and Hz band (ET~\cite{Punturo:2010zz} and CE~\cite{Evans:2021gyd}),
(III) MHz and higher band.
  Right: SGWB from stellar-mass BH, IMBH, SMBH and primordial BH (PBH) mergers. For SMBHs the estimated event rate density is $ 10^{-5}-10^{-2}~{\rm Gpc}^{-3}{\rm yr}^{-1}$~\cite{2021MNRAS.502L..99M,2024A&A...685A..94E}, for StMBHs it is $ 10-100~{\rm Gpc}^{-3}{\rm yr}^{-1}$~\cite{theligoscientific}. For IMBHs it is very uncertain and can be between $10^{-2}$ to 10 ${\rm Gpc}^{-3}{\rm yr}^{-1}$~\cite{2022ApJ...933..170F}. 
  For PBHs, we take $10^{-4} < \rho_{\rm PBH} / \rho_{\rm DM}
 < 10^{-3}$ and $M_{\rm PBH} \sim 10^{-6} {\rm M}_\odot$ as a fiducial value.
  }
  \label{fig:singlephenomultiplefreq}
\end{figure*}

In radiation domination we have $\Omega_{\rm GW, mem} \propto f^3\,\ln^2 f$ and $\Omega_{\rm GW, lin} \propto f^3 $; in matter domination we have $\Omega_{\rm GW, mem} \propto f$ and $\Omega_{\rm GW, lin} \propto f $; 
in kination domination we have $\Omega_{\rm GW, mem} \propto f^2$ and $\Omega_{\rm GW, lin} \propto f^4 $. They are also shown in the left panel of Figure \ref{fig:singlephenomultiplefreq}.

{\it Shape of the Spectrum and Transition to Memory Domination in the Infrared---}
For stochastic GW background (SGWB) from binary black hole (BBH) mergers, there is no linear GW signal at frequencies below maximal separation distance of heaviest black holes (BH). For frequencies smaller than this high-mass turn-around, the linear GW scales with $f^3$, therefore the GW spectrum of memory surpasses the linear signal in the infrared spectrum as shown in the right panel of Figure \ref{fig:decomposenonlinearmemory}. 
For astrophysical sources, the linear signal scales with $f^{2/3}$ from peak frequency ($f_{\rm peak}$) to turn-around frequency ($f_{\rm ta}$) and scales with $f^3$ for frequencies lower than turn-around, i.e. 
$\Omega_{\rm GW,lin} \propto \left( \frac{f}{f_{\rm peak}} \right)^{2/3} \left( \frac{f}{f + f_{\rm ta}} \right)^{7/3} $.
We find the frequency at which memory domination starts as
\begin{multline}
    f_{\rm mem. \, dom.}^{\rm astro} \sim f_{\rm ta} \; \left( \frac{f_{\rm ta}}{f_{\rm peak}}\right)^{1/6} \sqrt{\xi} \\
    \sim 10^{-6.5} f_{\rm peak} 
\, \left( \frac{f_{\rm ta}/f_{\rm peak}}{10^{-4}} \right)^{7/6} \sqrt{\frac{\xi}{3\cdot 10^{-4}}} 
\label{eqastromemdom}
\end{multline}
$f_{\rm peak}$ is near kHz for stellar-mass BH (StMBH), Hz for intermediate mass BHs (IMBH), and $\mu$Hz for supermassive BHs (SMBH). $f_{\rm ta}$ is 3-5 decades lower than corresponding peak frequency, and $
 \xi \equiv \Omega_{\rm GW,mem} / \Omega_{\rm GW,lin}  \approx 2-3\cdot 10^{-4}$ is the linear to memory ratio at peak which is independent of mass and event rate and it is approximately same for all BBH mergers up to a factor of a few depending on spins and eccentricities.

For cosmological sources, if we have $n>1$, corresponding to $-1/3<w<1/3$, then both linear and memory signals obeys causality tail scaling, hence memory is subdominant to linear signal at all frequencies. For $n\leq1$, corresponding to $w\geq 1/3$ (stiffer equation of states), the linear signal fades away faster than memory and memory dominates low frequency regimes.  We find the frequency at which memory domination starts, by assuming linear signal scales with $f^3$ in the low frequency regime and universe is in radiation domination, as follows
\begin{eqnarray}
  f_{\rm mem. \, dom.}^{\rm cosmo} &\sim& \frac{{\cal H}_*}{2\pi} \; {\rm exp}\left[{- \sqrt{\frac{1}{ \frac{k_*^2}{{\cal H^{\rm 2}_*}}\Delta_h^2(k_*)}}} \, \right]   \nonumber\\
  &\sim& 10^{-4.5} \frac{{\cal H}_*}{2\pi} \; {\rm exp}\left[- \frac{k_*}{{\cal H}_*} \; \frac{0.1}{\epsilon} \right] 
  \label{eqcosmomemdom}
\end{eqnarray}
where we used $\Delta_h^2\equiv \epsilon^2 \left( \frac{{\cal H}_*}{k_*} \right)^4$ in the last line, and $\epsilon$ being the dimensionless factor representing anisotropic energy-stress tensor, and ${\cal H}_*=H_* (a_*/a_0)$. We generalize the memory domination frequency for generic background with $n<1$ ($w>1/3$) 
\begin{eqnarray}
  f_{\rm mem. \, dom.}^{\rm cosmo} &\sim& \frac{{\cal H}_*}{2\pi} \left( \frac{k_*^2}{{\cal H^{\rm 2}_*}}\Delta_h^2(k_*) \right)^{\frac{1}{4(1-n)}} \nonumber\\
  &\sim& 0.1^{\frac{1}{4(1-n)}} \; \frac{{\cal H}_*}{2\pi} \; \left( \frac{{\cal H}_*}{k_*} \frac{\epsilon}{0.1} \right)^{\frac{1}{4(1-n)}} 
  \label{eqgenericcosmomemdom}
\end{eqnarray}

{\it Cosmological, Primordial and Astrophysical Sources---}
We calculate the amplitudes of the nonlinear memory signals of various cosmological sources, including SGWB from (p)reheating~\cite{Amin:2014eta,Caprini:2018mtu}, scalars~\cite{Domenech:2021ztg}, inflation~\cite{Guzzetti:2016mkm}, cosmic strings~\cite{Caprini:2018mtu}, phase transition~\cite{Caprini:2018mtu}, domain walls~\cite{Hiramatsu:2013qaa}, turbulence and magnetic fields~\cite{Caprini:2015zlo},
which are summarized in Table \ref{tab:summary}. In the low-frequency regime, memory signal becomes comparable or more dominant depending on the physical parameters of the system such as $\Delta_h^2(k_*)$ and $k_*/{\cal H}_*$ given as in Eq. \eqref{eqomegagwforgenericeos}. In Figure \ref{fig:singlephenomultiplefreq}, we show 3 cosmic SGWBs, which are labeled as (I), (II) and (III) with peaks at different frequencies, i.e. at different energy scales. Solid (dashed) lines show the total (nonlinear memory) signal. These SGWBs are produced under different equations of state, therefore their memory and causal scalings are different. For $n\leq1$ ($w>1/3$), as in the case of radiation domination (I) and kination domination (III), memory dominates the low frequency spectrum. For $n>1$ ($-1/3<w<1/3$) the causal tail of the linear signal dominates the infrared.

\begin{table}
\centering
\begin{tabular}{|l|l|l|}
  \hline
{\bf SGWB types}     & {\bf $\Omega_{\rm GW, lin}$} & {\bf $\Omega_{\rm GW, mem}$} \\  \hline
(P)Reheating  &   $10^{-7}$      &  $10^{-9}$ \\   \hline
Scalar Induced   &   $10^{-9}$      &  $10^{-13}$  \\  \hline
Inflation    &   $10^{-9}$      &  $10^{-13}$ \\   \hline
Cosmic Strings   &   $10^{-7}$      &  $10^{-9}$ \\  \hline
Phase Transitions  &   $10^{-8}$      &  $10^{-11}$ \\  \hline
Domain Walls    &   $10^{-8}$      &  $10^{-11}$ \\  \hline
Turbulence and Magnetic Fields \;\;\; \;\;\; \;\;
&   $10^{-8}$  &  $10^{-11}$ \\   \hline
Binary PBH  &   $10^{-10}$      &  $10^{-13}$ \\  \hline
Binary StMBH [$\lesssim \mathcal{O}(10^2)$~Hz]   &   $10^{-9}$      &  $10^{-13}$ \\  \hline
Binary NS / BHNS  [$\lesssim \mathcal{O}(10^3)$~Hz]   &   $10^{-9}$      &  $10^{-14}$ \\  \hline
Binary IMBH [$\lesssim\mathcal{O}(10^{-1})$~Hz]  &   $10^{-12}$      &  $10^{-15}$ \\  \hline
Binary SMBH [$\lesssim\mathcal{O}(10^{-5})$~Hz] &   $10^{-9}$      &  $10^{-13}$ \\  \hline
  \end{tabular}
  \caption{Summary of the SGWB energy density peak from the linear emission and the nonlinear memory.}
  \label{tab:summary}
  \vspace{-0.5cm}
\end{table}

The main astrophysical sources contributing to the stochastic nonlinear memory background are compact binary mergers (BBH, BNS, BBHNS), and among them, BBH mergers are expected to be the leading source. Using the merger estimates for the stellar-mass, intermediate mass, supermassive BBH populations, we calculate the linear GW and nonlinear GW memory energy density. We did the same for hypothetical primordial BH (PBH) mergers. The results are shown on the right panel in Figure \ref{fig:singlephenomultiplefreq}  and listed in Table \ref{tab:summary}.

{\it Observability of the nonlinear GW memory background---}\,
Direct detection of the nonlinear memory background requires it to be the dominant stochastic signal. Competition arises primarily from the linear component, which is overcome at deep infrared frequencies since the nonlinear memory decays slower. However, the signal amplitude in this dominance regime may be too weak for reliable measurement. Additional competition stems from other stochastic astrophysical backgrounds once the linear signal is surpassed. These challenges are overcome through source resolution. Next-generation, gravitational-wave detectors such as ET and CE are projected to resolve essentially all linear compact binary merger signals across the cosmos. This strategy effectively removes the linear signal background, and isolates the nonlinear memory and residual noise from parameter estimation uncertainty. Similarly, future observatories LISA-TianQin-Taiji and SKA can also resolve the individual IMBH and SMBH sources to extract memory component. For cosmological sources, the nonlinear memory can be detected using its unique properties, such as quadratic logarithmic growth, running slope, and induced polarization.

{\it Discussion and Conclusions---}
We calculated the universal nonlinear memory contribution to the SGWB. We found that for wavelengths shorter than the cosmological horizon at the time of formation, the spectrum scales universally with $f$, whereas for longer wavelengths it scales with $f^3\ln^2 (1/f)$ in the radiation domination era. We generalized our scaling relations and find that memory scales with $f^{3 - 2 \vert \frac{3w-1}{3w+1} \vert}$ for a generic equation of state of the universe. Compared to the universal $f^3$ scaling of the transient sources, memory becomes the dominant component of the infrared SGWB.

We calculated nonlinear memory backgrounds for different cosmological and astrophysical sources. There are many possibilities for cosmological sources with a wide range of amplitude and frequency ranges. The main astrophysical sources are the BBH mergers in different mass ranges. Nonlinear memory dominates over linear signal in the low frequency regime, which allows us to derive complete frequency spectrum of astrophysical SGWB. Future multi-band gravitational wave detectors can enable the observation of the astrophysical nonlinear memory background via its low frequency universal linear scaling or resolving the linear sources. The cosmological memory background will similarly be accessible via its distinct spectral modifications.

{\it Acknowledgements---}
We are grateful to Bayram Tekin, David Garfinkle, David Maibach, Antonio Riotto, and Töre D. Boybeyi for the discussions. We further thank Ahmet Bingül, Seçkin Kürkçüoğlu, Fethi M. Ramazanoğlu, and Stephen R. Taylor for the suggestions on the manuscript.
\\
C.\"U. and D.V. are supported by TÜBİTAK (The Scientific and Technological Research Council of Türkiye) through the grant number 123C484. C.\"U. acknowledges support from EU Cost Actions CA21136 (CosmoVerse) and CA21106 (Cosmic Wispers), and the METU DOSAP-C program. D.V. acknowledges support from the European Union’s Horizon Europe Research and Innovation Programme under the Marie Skłodowska-Curie Actions (MSCA) COFUND Programme with the grant number 101081645 and TÜBİTAK with the grant number 123C213.
\\
During the completion of this work, C.\"U. thanks for the hospitality of Flatiron Institute Center for Computational Astrophysics, Harvard \& Smithsonian Center for Astrophysics, Heidelberg University, METU Research and Application Center for Space and Accelerator Technologies (METU IVMER) and New York University; also for the support, hospitality and companionship of Özlü Aran, Ahmet Bingül, Bilge Demirköz, Şakir Erkoç, Mürsel Karadaş, Hasan Karpuz, Ertan Kuntman, Seçkin Kürkçüoğlu, Sarah Libanore, Osman Barış Malcıoğlu, Anna Maslarova, Federico Oliva, Kaan Sen, Erica Pacchioni and Yiyao Zhang.
\\
C.\"U. dedicates this work to Mahmut Çağlıyurt, Nevgün Çağlıyurt, Hamiyet Ünal and Ömer Ünal.

\newpage
\bibliographystyle{apsrev4-1}
\bibliography{main.bib}

\newpage
\section{End Matter}
\subsection{Analytic Evaluation of Nonlinear Memory from Scaling of Linear Spectrum}
To get analytical results, we may further assume a power spectrum with the form 
\begin{equation}  \label{eqphform}
  \Delta_h^2 (k) \simeq \frac{2 \Delta_h^2 (k_*)}{\left(\frac{k_*}{k}\right)^a + \left(\frac{k}{k_*}\right)^b} 
\end{equation}
peaking around $k_*$, where $\Delta_h^2 (k) \propto \Delta_h^2 (k_*) \left(\frac{k}{k_*}\right)^a$ in the $k<k_*$ regime, and $\Delta_h^2 (k) \propto \Delta_h^2 (k_*) \left(\frac{k}{k_*}\right)^{-b}$ in the $k>k_*$ regime. The low frequency slope is set by causality or late decaying sources as $a\simeq 2-3$ and high frequency slope $b>2$ is set by the nature of the process. Then we have \footnote{Note that we can also describe the peak of the spectrum in terms of Gaussian shape, in that case again, the result will depend on with and how far this functional form extends in describing the spectrum.}
\begin{equation}
  {\cal F}_k \equiv \frac{x^{2+b}}{2+b} {\rm 2\, HypergeoF\, 1} \left[1, \frac{2+b}{a+b}, 1+\frac{2+b}{a+b},-x^{a+b}\right]
\end{equation}
where $k_b$ is maximum momentum with slope $b$, larger than $k_*$, and $k_a$ is minimum momentum with slope $a$, smaller than $k_*$, where the power spectrum form given in Eq. \eqref{eqphform} reasonably holds. As long as $k_b \sim O(1) k_*$ and $k_a \sim \frac{k_*}{O(1)}$, then $ {\cal F}_k (k_b,k_a) ={\cal F}_k (k_b) - {\cal F}_k (k_a) \sim O(1)$.

\subsection{Green Functions for Generic Equation of State}
Green functions in distinct equations of state is given by
\begin{equation}
  G_k(\tau,\tau') = \frac{U_1(k \, \tau) U_2 (k \, \tau') - U_1(k \, \tau') U_2 (k \, \tau)}{{\cal N}} 
\end{equation}
where ${\cal N} = W_{U_1, U_2} = \frac{2}{\pi} k$ is the Wronskian of $U_1 = \sqrt{x} J_{n-1/2}(x)$ and $U_2= \sqrt{x} Y_{n-1/2}(x)$ are homogeneous solutions of GW equation in distinct equations of states, Bessel functions. 
In general, we have 
\begin{equation}
 \!\!\! \!\! G_k (\tau,\tau')= 
 \frac{\pi}{2k} \sqrt{x \; x'} (J_{n-\frac{1}{2}}(x)Y_{n-\frac{1}{2}}(x')- J_{n-\frac{1}{2}}(x')Y_{n-\frac{1}{2}}(x))
 \label{eqgenericgreen}
\end{equation}

In the deep horizon Green function is independent of the equation of state, i.e. the expansion rate of the universe and match with local Minkowski result.
\begin{equation}
  G_k(\tau,\tau') = \frac{\sin [k (\tau - \tau')]}{k} \nonumber
\end{equation}

\subsection{Derivation of Analytic Nonlinear Gravitational-Wave Memory Background Given in Equations \eqref{eq:genericformemory} and \eqref{eqomegagwforgenericeos}
}
\label{secdetailsofapproximation}
Here we put the details leading to the derivation of our result in Eq. \eqref{eq:genericformemory}. 
\begin{figure}
  \centering
  \includegraphics[width=\linewidth]{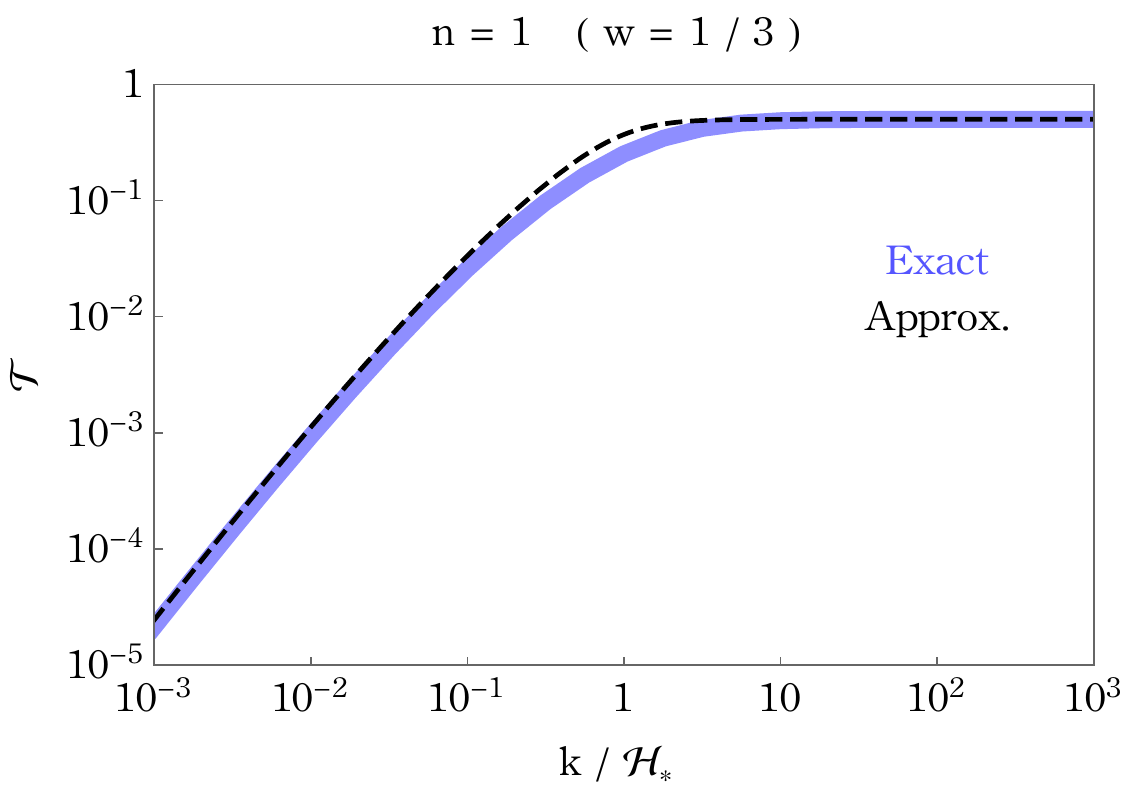}
  \caption{Comparing exact time integral and our approximation in Eq. \eqref{eq:timeintegapproximation} which leads to final expression \eqref{eq:genericformemory}
  }
\label{fig:approximationN1}
\end{figure}

Time integral is in the form
\begin{multline}
\!\! \!\! \!\! {\cal I} (x,x_*)
  = \int_{\eta_*}^{\eta} \frac{d \eta'}{\eta'/\eta_*} \frac{\sin k (\eta - \eta')}{k} = \frac{x_*}{k^2} \int_{x_*}^x \frac{d x'}{x'} \sin (x-x') \\
  = \frac{x_*^2}{k^4} \bigg( \sin x \left( Ci(x)-Ci(x_*) \right) - \cos(x) \left( Si(x) - Si(x_*)\right) \bigg) 
\end{multline}
where $x=k\eta$, $x'=k\eta'$, $x_*=k\eta_* = k/{\cal H}_*$, $Ci$ is Cosine Integral, $\int_0^x \frac{\cos y}{y} dy$, and $Si$ is SineIntegral, $\int_0^x \frac{\sin y}{y} dy$.

After imposing momentum conservation $\delta^3 ({\bf k} + {\bf k'})$, we have $({\bf k} + {\bf k'})$ and the square of this time integral for the final expression. This final quantity needs to be averaged over time for multiple periods in the limit $x, x_1, x_2 \gg 1$. 
\begin{eqnarray}
  {\cal T} (x) &\equiv& k^2 \; \overline{ {\cal I}(x,x_*)^2} = \frac{k^2}{x_2 - x_1} \int_{x_1}^{x_2} dx \; {\cal I}(x,x_*)^2 \nonumber\\
  &\simeq& \frac{1}{2} \left(1 + \ln^2{ \left(\frac{x_*}{1+x_*} \right)} \right) \left(\frac{x_*}{1+x_*} \right)^2
  \label{eq:timeintegapproximation}
\end{eqnarray}
where $\Omega _{\rm GW} \propto k^5 \frac{{\cal T}}{k^4} \propto k {\cal T}$.
We compare exact result and our approximation in Figure \ref{fig:approximationN1}. For radiation domination era, the difference between analytical approximation and exact result is smaller than 1 \% for $x_* \ll1$, around 10\% for $x_* \ll 1$, and around 10-30 \% for $x_* \sim {\cal O}(1)$.

\begin{figure}
  \centering
  \includegraphics[width=\linewidth]{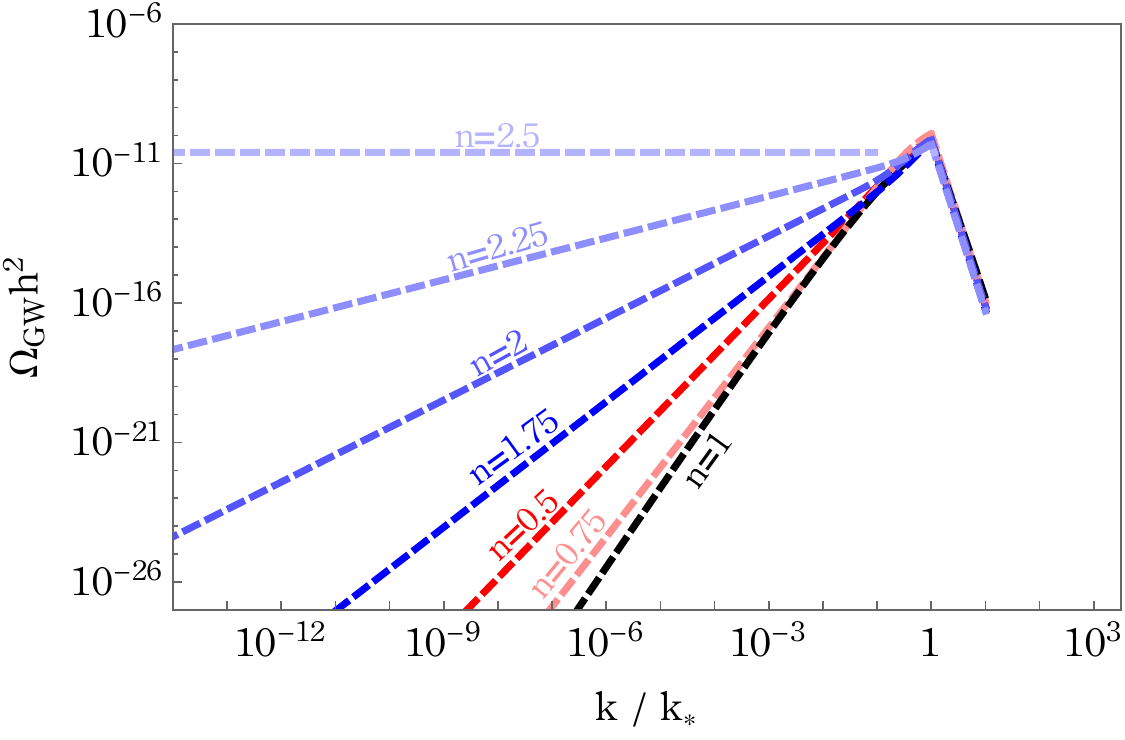}
  \caption{Memory slope with distinct equation of states ($n=\frac{2}{3w+1}$) for $k_*\sim {\cal H}_*$. Red curves ($n<1$) due to persistent source and blue curves ($n>1$) due to causal slope. $n=0.5, \, 0.75, \, 1,\, 1.75,\, 2,\, 2.25, \, 2.5$ correspond to $w = 1,\, 5/9, \, 1/3, \, 1/21,/, 0, \, -1/27, \, -1/15$, respectively.
  }
\label{figmemoryslopewitheos}
\end{figure}

\begin{figure}
  \centering
  \includegraphics[width=0.71\linewidth]{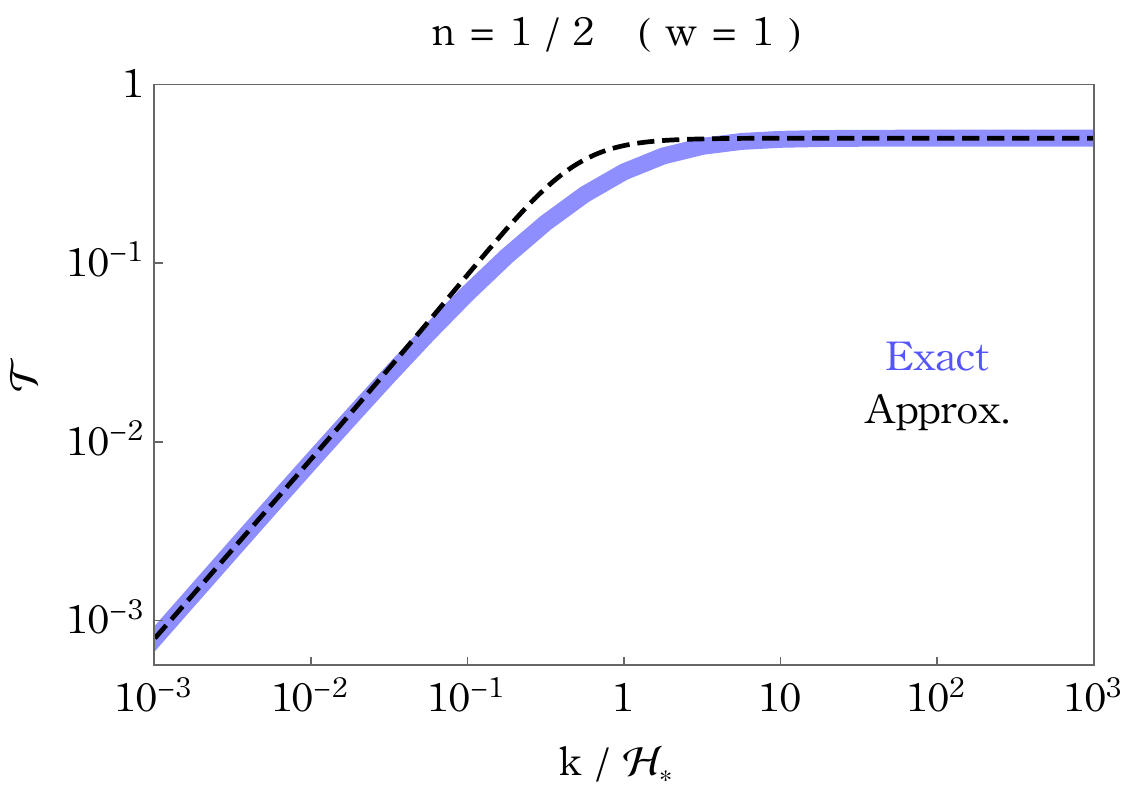}
   \includegraphics[width=0.71\linewidth]{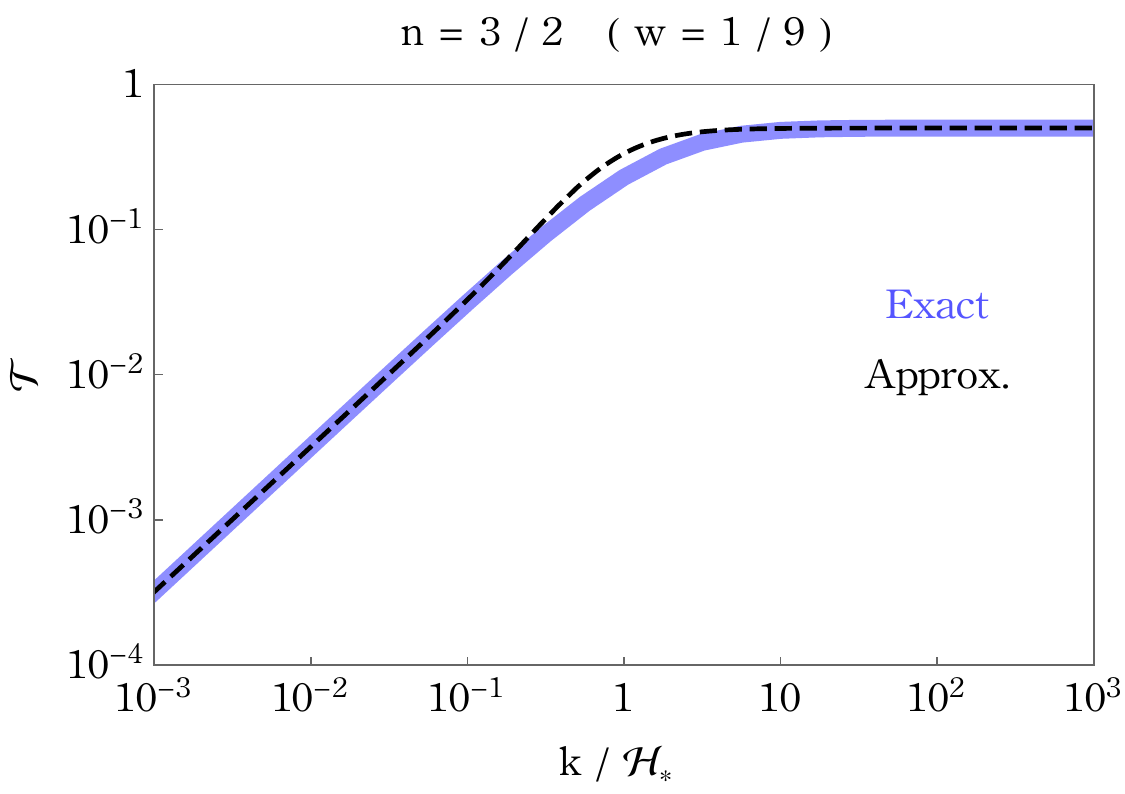}
    \includegraphics[width=0.71\linewidth]{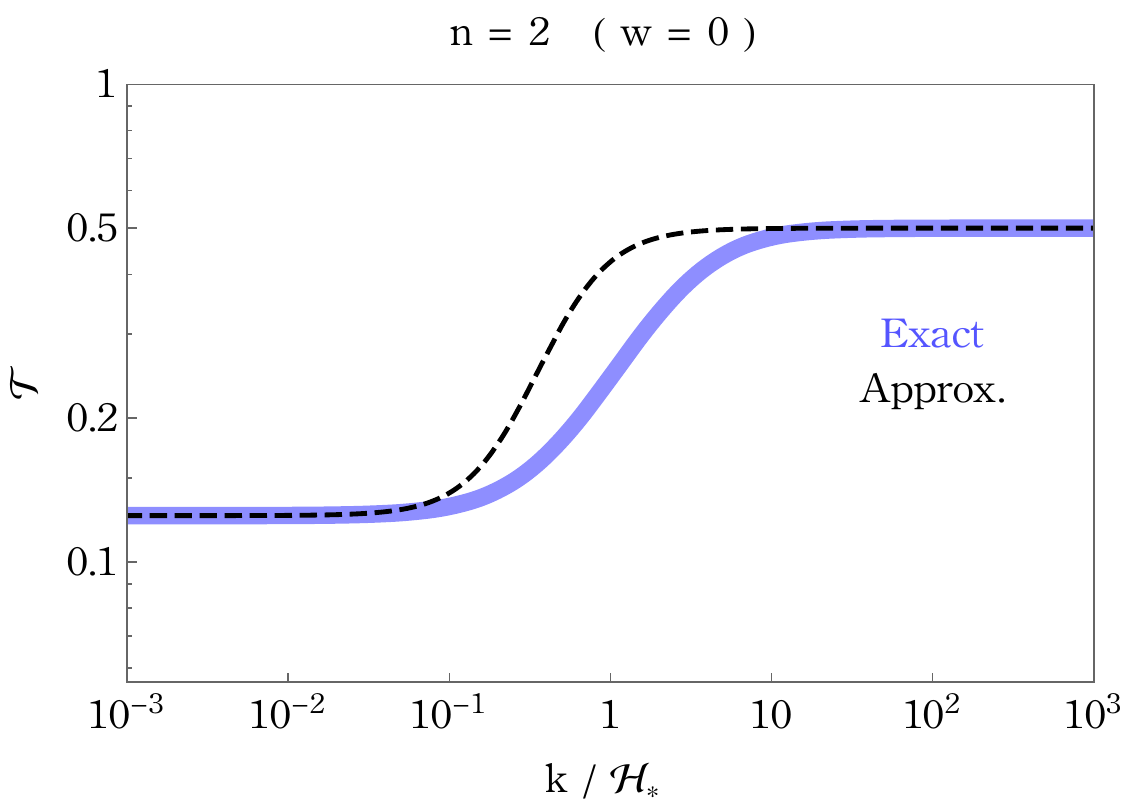}
    \includegraphics[width=0.71\linewidth]{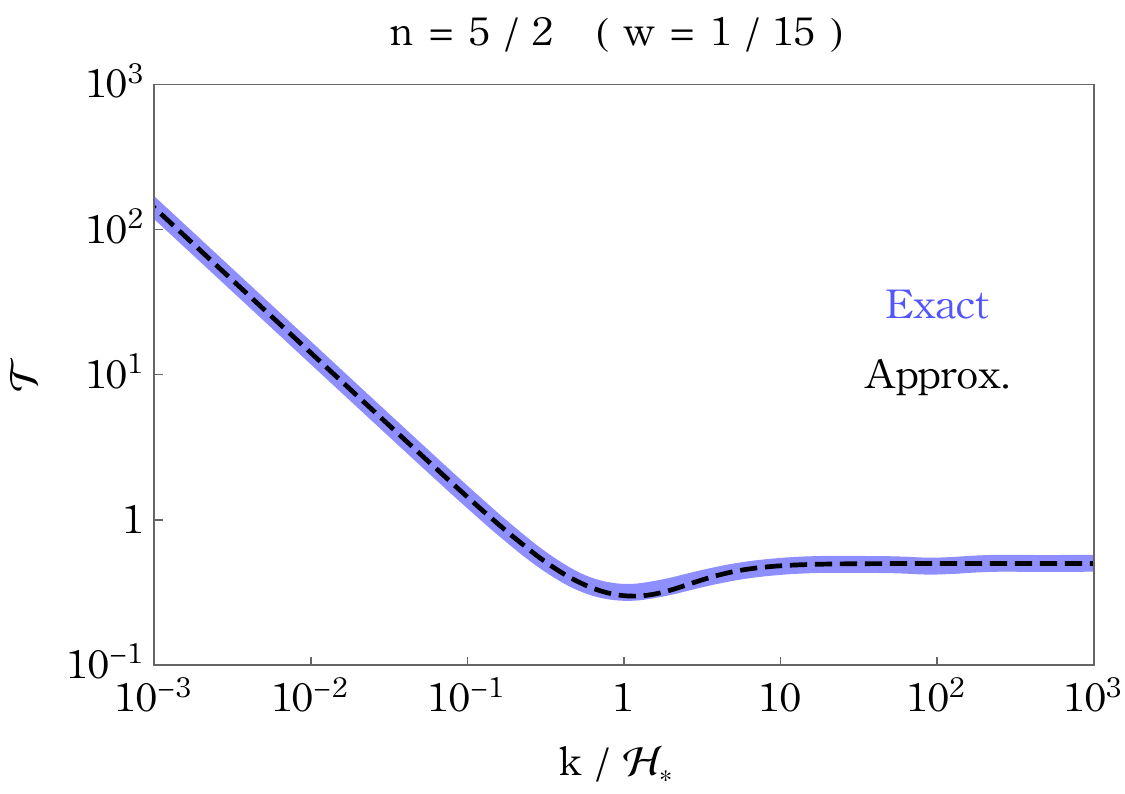}
    \includegraphics[width=0.71\linewidth]{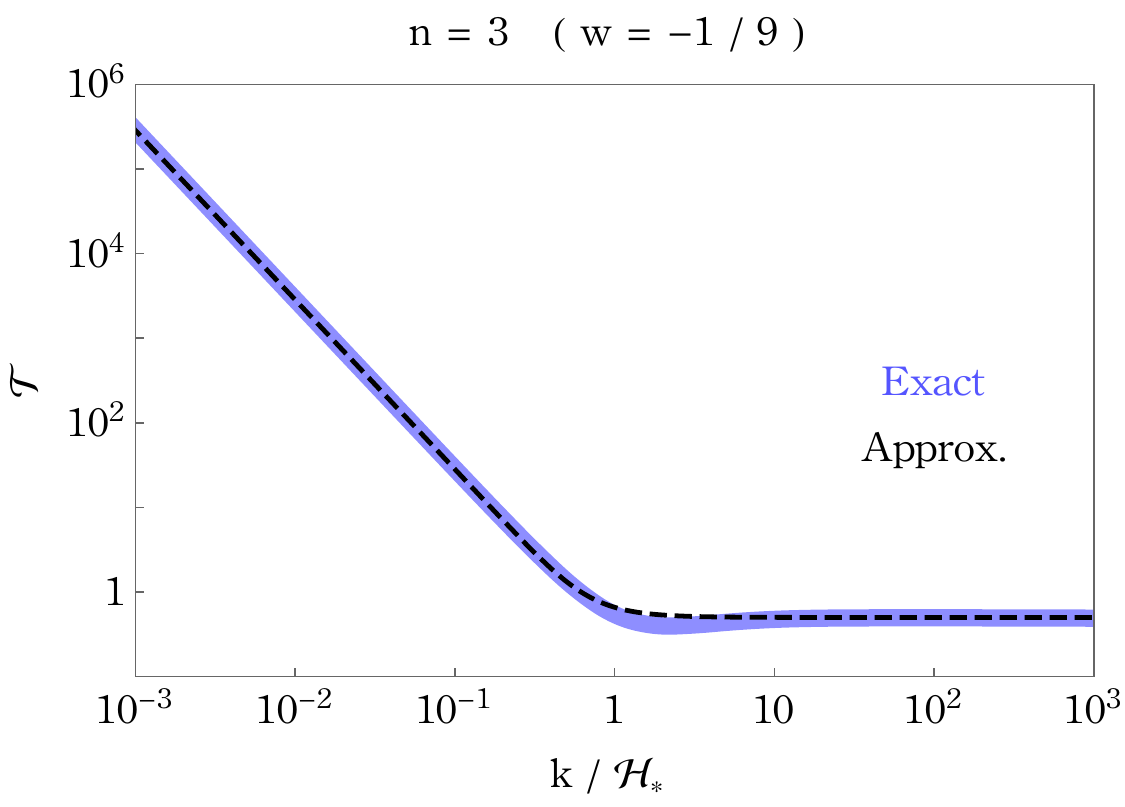}
  \caption{Comparison of the exact and approximate solutions for $n={0.5, \, 1.5, \, 2, \, 2.5, \, 3}$ corresponding to equations of states $w={1, \, 1/9, \, 0, \, 1/15, \, -1/9}$, respectively.}
\label{fig:approximationgenericN}
\end{figure}

We further generalize our results for an arbitrary equation of state \eqref{eqomegagwforgenericeos}. We numerically solve nonlinear gravitational-wave (GW) memory for generic equation of state with generic Green function given in Eq. \eqref{eqgenericgreen}. There are two branches of solutions for the generic expansion, namely if the expansion is fast, i.e. case for $w<1/3$, such that earlier superhorizon then subhorizon GWs dilute away rapidly, then there will be no persistent (non-decaying) sourcing and sourcing is temporary and infrared scaling is set by the causality tail which scales as $\Omega _{\rm GW} \propto f^{5-2n}$. However, if the there is stiffer equation of state, i.e. $w>1/3$, then GW sourcing behaves like a persistent source and the nonlinear memory dominates over causality tail. As a result, GW background scales as $\Omega _{\rm GW} \propto f^{1+2n}$ in the infrared regime. Combining $x_*\gg1$ and $x_*\ll1$ limits, we extract out the GW memory signal behavior and find reasonably good approximation to exact solution in terms of polynomials, which allows us to read all qualitative properties of the background. The approximate solution for a generic equation of state becomes
\begin{multline}
   \Omega_{\rm GW, \, mem} \simeq {\cal A} \; \frac{k}{k_*} \; \frac{ (k \eta_*)^2}{1+ 4 (k \eta_*)^2} \\
   \times \left[ \left( \frac{ (k \eta_*)^2}{{\cal C}_1^{\frac{1}{1-n}}+ (k \eta_*)^2} \right)^{n-1} + \left( \frac{ (k \eta_*)^2}{{\cal C}_2^{\frac{1}{n-1}}+ (k \eta_*)^2} \right)^{1-n} \right]
\end{multline}
where 
\begin{eqnarray}
  {\cal C}_1 &=&\frac{2^{-2n-1} \; \pi^2 \; \csc[n \pi]^2}{\Gamma[n]^2}, \nonumber\\
  {\cal C}_2 &=& \frac{2^{2n-5}\; \pi \; \sec[n \pi]^2 \; \Gamma(1-n)^2}{\Gamma[3/2 - n]^2 \;\, \Gamma[2-n]^2},
\end{eqnarray}
valid for $w\neq1/3$, and for $w=1/3$ see Eq. \eqref{eq:genericformemory}. We verify that our analytic approximation is accurate up to ${\cal O}(10^{-5}-1)$ for various of equation of states, including $n={0.5, \, 1.5,\, 2, \, 2.5, \, 3}$ corresponding to $w={1, \, 1/9, \, 0, \, 1/15, \, -1/9}$, shown in Figure \ref{fig:approximationgenericN}.

\newpage

\end{document}